\input{aipcheck}

\documentclass[
    ,final            
  ]
  {aipproc}

\layoutstyle{6x9}

\usepackage{graphicx}
\usepackage{psfrag}
\newcommand{\be}{\begin{eqnarray}}
\newcommand{\ee}{\end{eqnarray}}

\newcommand{\ket}[1]{\vert\,{#1}\rangle}

\begin{document}

\title{Chiral Odd Generalized Parton Distributions in Position Space}


\author{A. Mukherjee}{
  address={Department of Physics,
Indian Institute of Technology, Powai, Mumbai 400076,
India.}
}
\author{D. Chakrabarti}{
  address={Department of Physics,
Swansea University, Singleton Park, Swansea, SA2 8PP, UK}
}
\author{R. Manohar}{
  address={Department of Physics,
Indian Institute of Technology, Powai, Mumbai 400076,
India.}
}

\begin{abstract}
We report on a calculation of the chiral odd generalized parton distributions 
(GPDs) for non-zero skewness $\zeta$ in transverse and longitudinal position
spaces by taking Fourier transform with respect to the transverse and
longitudinal momentum transfer respectively using overlaps of light-front
wave functions (LFWFs). 
\end{abstract}

\maketitle


\section{Introduction}
At leading twist, there are three forward parton distributions
(pdfs), namely, the unpolarized, helicity and transversity
distributions. Similarly, three leading twist generalized quark
distributions can be defined which in the forward limit, reduce to
these three forward pdfs. The third one is chiral odd and is
called the generalized transversity distribution $F_T$. This is defined as
the off-forward matrix element of the bilocal tensor charge operator.
It is parametrized in terms of four GPDs, namely $H_T$, $\tilde H_T$,
$E_T$ and $\tilde E_T$ in  the most general way \cite{markus,chiral,burchi}.
The chiral-odd GPDs affect the
transversely polarized quark distribution both in unpolarized and
in transversely polarized nucleon in various ways. A relation for the
transverse total angular momentum of the quarks has been proposed
in \cite{burchi}, in analogy with Ji's relation, which involves a
combination of second moments of $H_T, E_T$ and $\tilde H_T$ in the forward
limit. 

In a previous work we have investigated the chiral odd GPDs for a simple
spin-$1/2$ composite particle for $\zeta=0$ in impact parameter space
\cite{harleen}. In this work, using an overlap
formula in the terms of the LFWFs both in the
DGLAP ($n \rightarrow n$) and ERBL ( $n+1 \rightarrow n-1$) regions
\cite{ravi}, we investigate them in a simple model, namely  for
the quantum fluctuations  of a  lepton in QED at
one-loop order \cite{drell}. We generalize this analysis
by assigning a mass $M$ to the external electrons and a different
mass $m$ to the internal electron lines and a mass $\lambda$ to the
internal photon lines with $M < m + \lambda$ for stability.
In effect, we shall represent a spin-${1\over 2}$ system as a composite
of a spin-${1\over 2}$ fermion and a spin-$1$ vector boson
\cite{hadron_optics, dis,dip,dip2,marc}. This field theory inspired model
has the correct correlation between the Fock components of the state
as governed by the light-front
eigenvalue equation, something that is extremely difficult to achieve in
phenomenological models. GPDs in this model satisfy
general properties like polynomiality and positivity. So it is interesting
to investigate the general properties of GPDs in this model.
By taking Fourier transform (FT) with respect to $\Delta_\perp$, we express
the GPDs in transverse position space and by taking a FT with respect 
to $\zeta$ we express them in longitudinal position space.

\subsection{Chiral Odd GPDs using Overlap Formula}

We use the parametrization of \cite{burchi} for the chiral odd GPDs.
Following \cite{drell, hadron_optics}, we take a simple composite spin
$1/2$ state, namely an electron in QED at one loop to investigate the GPDs.
The light-front Fock state wavefunctions corresponding to the
quantum fluctuations of a physical electron can be systematically
evaluated in QED perturbation theory. The state is expanded in Fock
space and there are contributions from $\ket{e^- \gamma}$ and 
$\ket{e^- e^- e^+}$, in addition to renormalizing the one-electron state. 
Both the two- and three-particle Fock state components are given 
in \cite{overlap}. In the domain $\zeta <x <1$, there are diagonal $2 \to 2$
overlaps. Using the overlap formula in \cite{ravi} we calculate the chiral
odd GPDs in this kinematical region. They are given by;

\begin{figure}
\centering
\includegraphics[width=4.5cm,height=4.5cm,clip]{fig1b.eps}
\hspace{0.2cm}%
\includegraphics[width=5cm,height=5cm,clip]{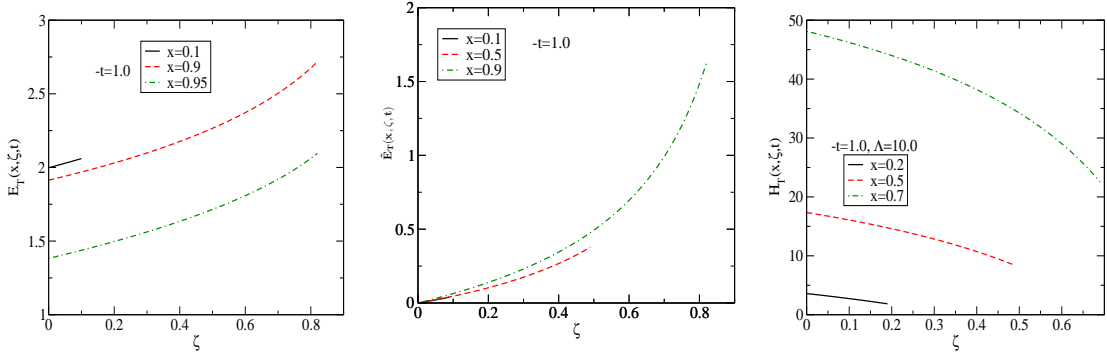}
\hspace{0.2cm}%
\includegraphics[width=4.5cm,height=4.5cm,clip]{fig3b.eps}
\caption{\label{fig1} (Color online) Chiral-odd
GPDs vs. $\zeta$  for fixed $t$ in  ${\mathrm{MeV}^2}$ and different values of
$x$.}
\end{figure}

\be
E_T(x,\zeta,t)&=&-{e^2\over 8 \pi^3}{2 M\pi\over 1-\zeta}(M-{m\over
x})x(1-x)
I_3,
\ee
\be
\tilde E_T(x,\zeta,t)&=&{e^2\over 8\pi^3}{M\pi\over 1-\zeta}\Big
[-(1-x)\Big\{
(M-{m\over x})x+(M-{m\over x'})x'\Big\}I_1\nonumber\\&&~~~~~~~~~
+(M-{m\over x})x(1-x)I_2\Big ],
\ee
\be
H_T(x,\zeta,t)&=&{e^2\over 8\pi^3}{\pi\over 2}\Big [{x+x'\over 2(1-x)}
\ln({\Lambda^4 \over DD'})+\Big\{{x+x'\over 2(1-x)}B(x,\zeta)
+{\zeta M \over 1-\zeta}(M-{m\over x})x(1-x)\Big\}I_2 \nonumber\\
&&-{\zeta M\over 1-\zeta}\Big\{(M-{m\over x})x(1-\zeta)+
(M-{m\over x'})x'\Big\}(1-x')I_1\Big ],
\ee
\be
I_1 &=&\int_0^1 dy{1-y\over Q(y)},\\
I_2 &=&\int_0^1 dy{1\over Q(y)};
\ee
where $Q(y)=y(1-y)(1-x')^2 \Delta_\perp^2-y(M^2x(1-x)-m^2(1-x)-\lambda^2x)
-(1-y)(M^2x'(1-x')-m^2(1-x')-\lambda^2 x')$ and
\be
I_3=\int_0^1 dy {y\over Q(y)}.
\ee
$D=M^2 x (1-x) -m^2 (1-x) -\lambda^2 x$ and $D'=M^2 x' (1-x') -m^2
(1-x') -\lambda^2 x'$.

In order to regulate the ultraviolet
divergences we use a cutoff $\Lambda$ on the transverse momentum 
$k^\perp$. $\tilde H_T(x,\zeta,t)$ is zero in this model.

In Fig. \ref{fig1} we have shown the chiral odd GPDs as functions of $\zeta$
for fixed $t$ and different values of $x$. 

Introducing the Fourier conjugate  $b_\perp$ (impact parameter)
of the transverse momentum transfer $\Delta_\perp$, the GPDs can
be expressed in impact parameter space. Like the chiral even counterparts,
chiral odd GPDs as well have interesting interpretation in impact parameter
space \cite{burchi}.  In most experiments $\zeta$ is nonzero, and it is
of interest to investigate the chiral odd GPDs in $b_\perp$ space with
nonzero $\zeta$. The probability interpretation is no longer possible as now
the transverse positions of the initial and final protons are different as
there is a finite momentum transfer in the longitudinal direction. The GPDs
in impact parameter space probe partons at transverse position $\mid b_\perp
\mid $ with the initial and final proton shifted by an amount of order
$\zeta \mid b_\perp \mid$. Note that this is independent of $x$ and even when
GPDs are integrated over $x$ in an amplitude, this information is still
there \cite{markus2}. Thus the chiral odd GPDs in impact parameter space
gives the spin orbit correlations of partons in protons with their centers
shifted with respect to each other. 
\begin{figure}
\centering
\includegraphics[width=5cm,height=5cm,clip]{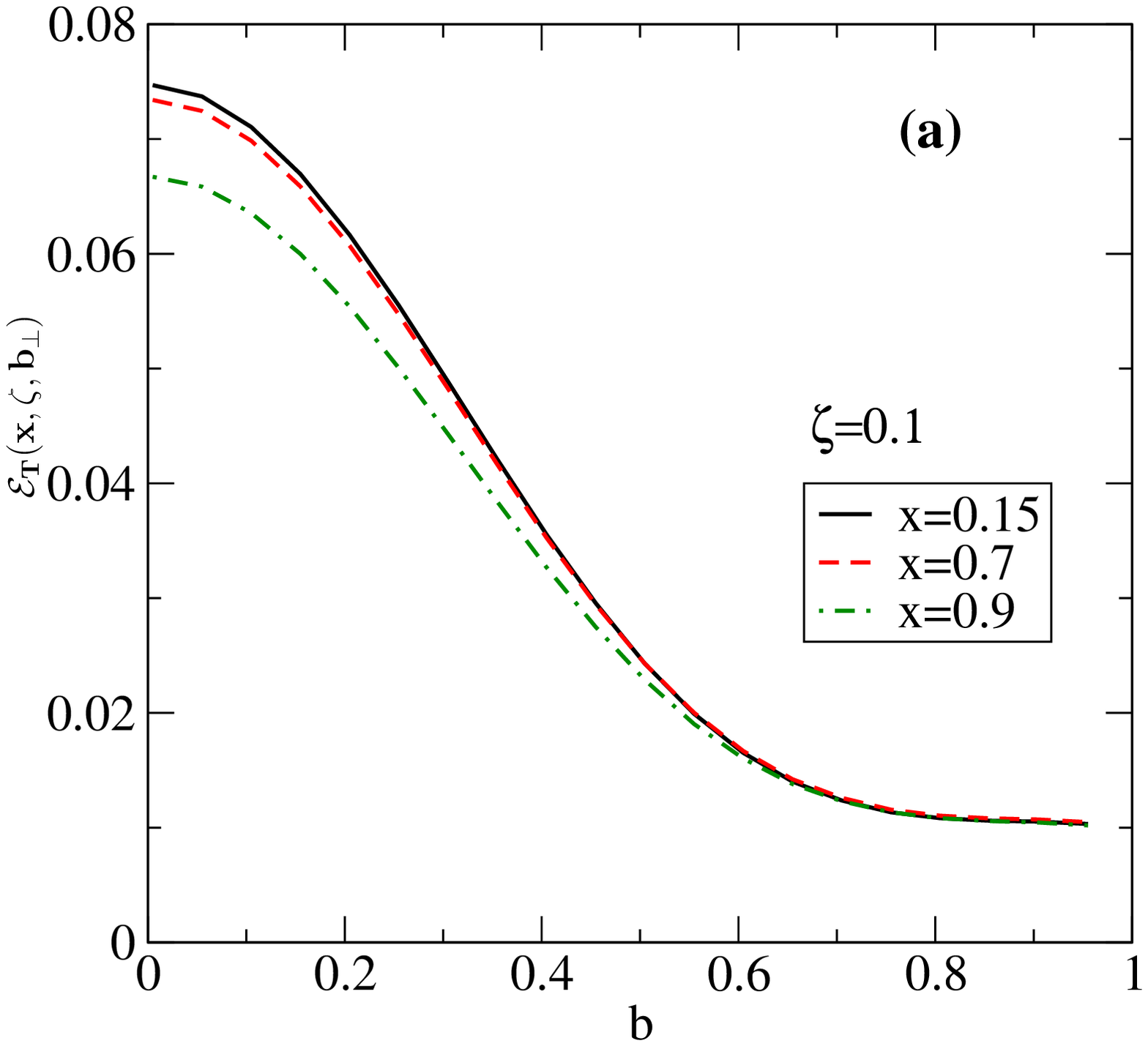}
\includegraphics[width=5cm,height=5cm,clip]{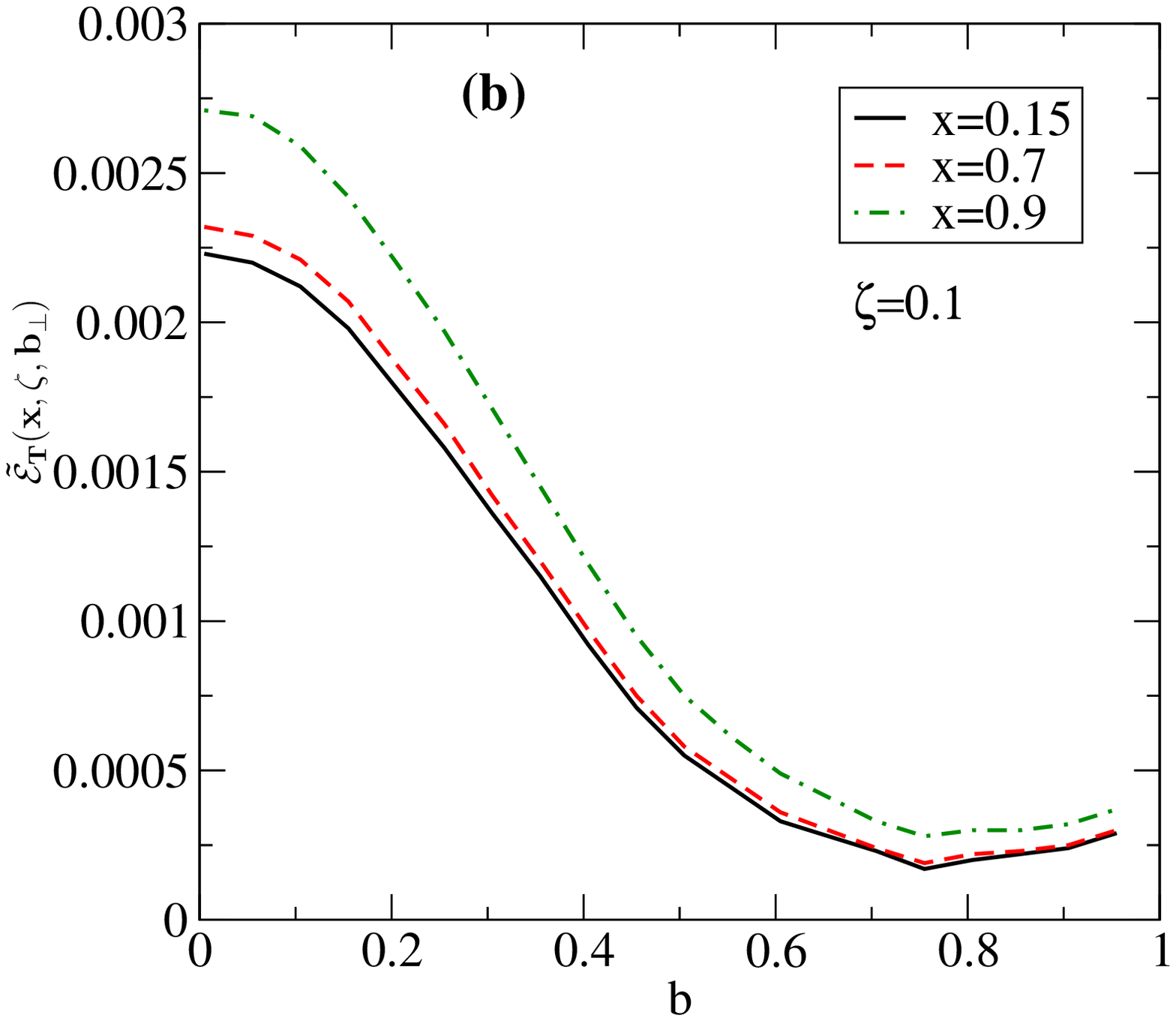}
\includegraphics[width=5cm,height=5cm,clip]{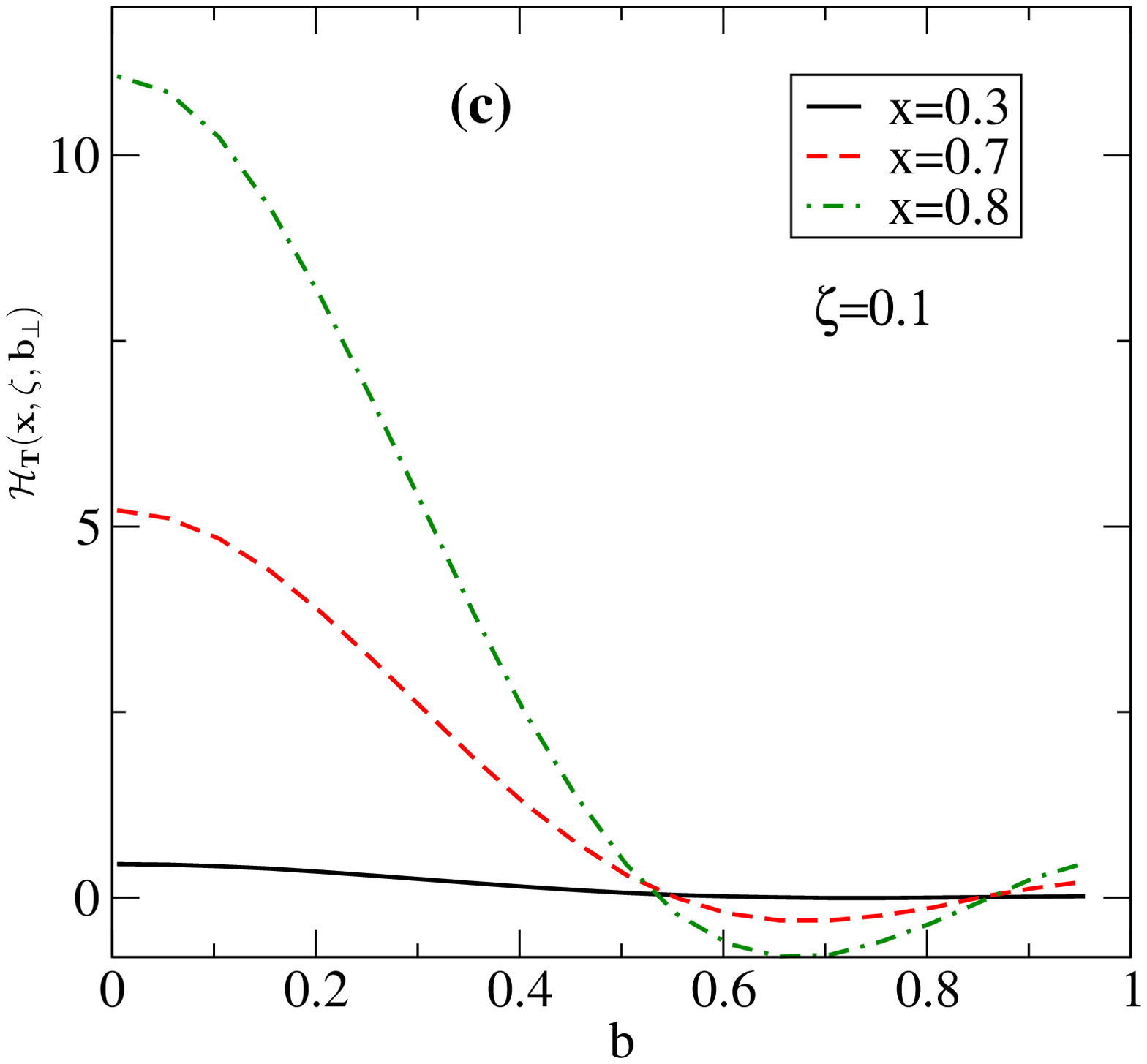}
\caption{\label{fig5} (Color online) Fourier spectrum of the chiral-odd
GPDs vs. $\mid b_\perp \mid $  for fixed $\zeta$  and
different values of $x$}
\end{figure}
Taking the Fourier transform (FT) with
respect to the transverse momentum transfer
$\Delta_\perp$ we get the GPDs in the transverse impact parameter space.
\be
{\cal E}_T(x,\zeta,b_\perp)&=&{1\over (2\pi)^2}\int d^2\Delta_\perp
e^{-i\Delta_\perp \cdot b_\perp} E_T(x,\zeta,t)\nonumber \\
&=&{1\over 2 \pi}\int \Delta d\Delta J_0(\Delta b) E_T(x,\zeta,t),
\ee
where $\Delta=|\Delta_\perp|$ and $b=|b_\perp|$. The other impact
parameter dependent GPDs $\tilde{\cal E}_T(x,\zeta,b_\perp)$ and
${\cal H}_T(x,\zeta,b_\perp)$ can also be defined in the same way.
It is to be noted that ${\cal H}_T(x,\zeta,b_\perp)$
for a free Dirac particle is expected to be a delta function; the smearing
in $\mid b_\perp \mid $ space is due to the spin correlation in the
two-particle LFWFs.
Fig. \ref{fig5} shows the plots of the above three functions
for fixed $\zeta$ and different values of $x$. For given $\zeta$, the peak
of ${\cal H}_T(x,\zeta,b_\perp)$ as well as $\tilde 
{\cal E}_T(x,\zeta,b_\perp)$
increases with increase of $x$, however for ${\cal E}_T(x,\zeta,b_\perp)$
it decreases.

In \cite{wigner}, a phase space distribution of quarks and gluons in the
proton is given in terms of the quantum mechanical Wigner distribution
$W(\vec{r}, \vec{p})$, in the rest frame of the proton, which are functions
of three position and three
momentum coordinates. Wigner distributions are not accessible in experiment.
However, if one integrates two momentum components one gets a reduced Wigner
distribution $W_\Gamma(\vec{r},x)$ which is related to the GPDs by a Fourier
transform.  For given $x$, this gives a 3D position space picture of the
partons inside the proton. If the probing wavelength is
comparable to or smaller than the Compton wavelength ${1\over M}$, where $M$
is the mass of the proton, electron-positron pairs will be created, as a
result, the static size of the system cannot be probed to a precision better
than  ${1\over M}$ in relativistic quantum theory. However, in light-front
theory, transverse boosts are Galilean boosts which do not involve dynamics.
So one can still express the GPDs in transverse position or impact parameter
space and this picture is not spoilt by relativistic corrections. However,
rotation involves dynamics here and rotational symmetry is lost. In
\cite{hadron_optics}, a longitudinal boost invariant impact parameter
$\sigma$ has been introduced which is conjugate to the
 longitudinal momentum transfer $\zeta$. It was shown that the DVCS
amplitude expressed in terms of the variables $\sigma, b_\perp$ show 
diffraction
pattern analogous to diffractive scattering of a wave in optics where the
distribution in $\sigma$ measures the physical size of the scattering
center in a 1-D system. In analogy with optics, it was concluded that the
finite size of the $\zeta$ integration of the FT acts as a slit of finite
width and produces the diffraction pattern.
\begin{figure}
\centering
\includegraphics[width=5cm,height=5cm,clip]{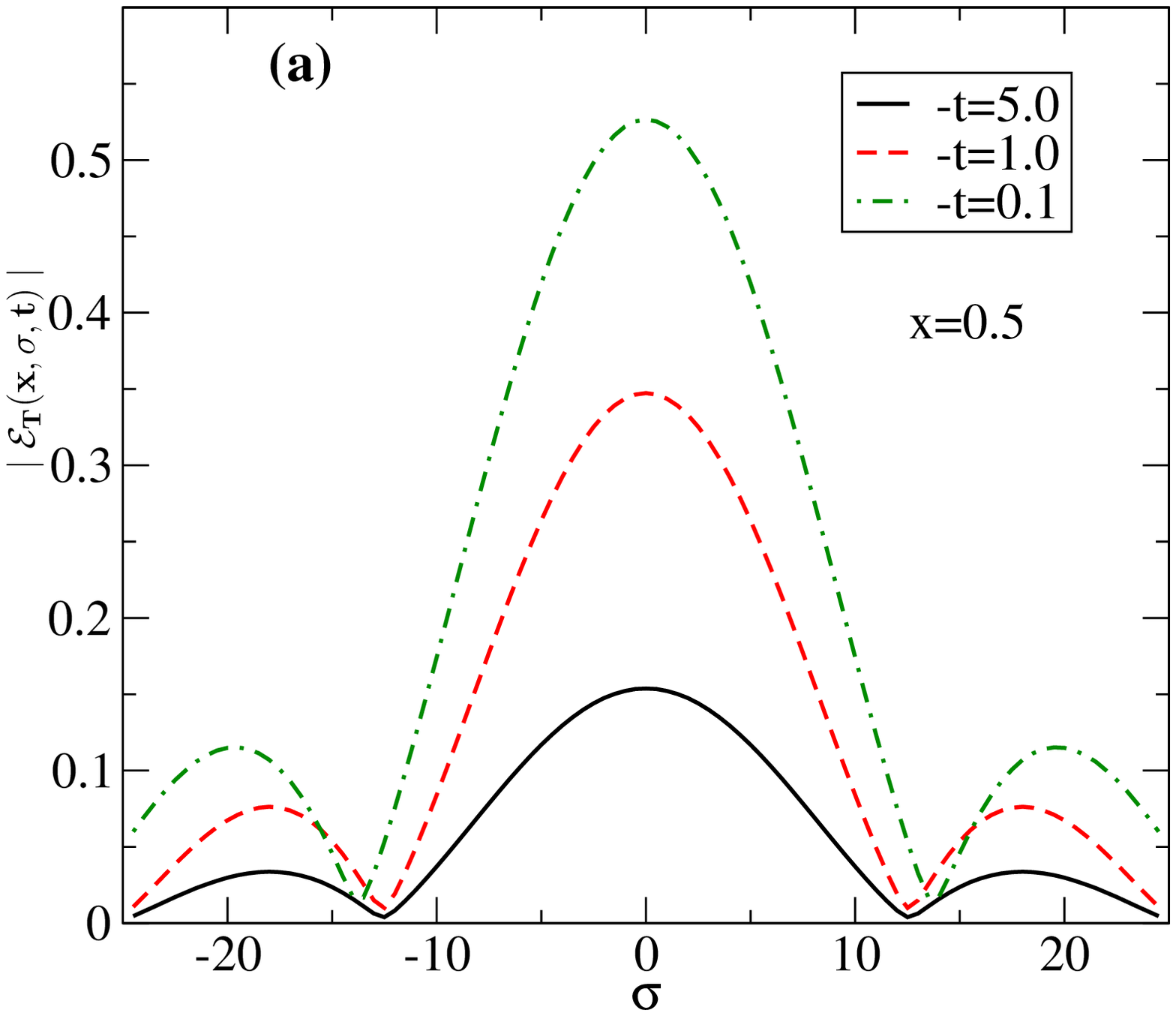}
\includegraphics[width=5cm,height=5cm,clip]{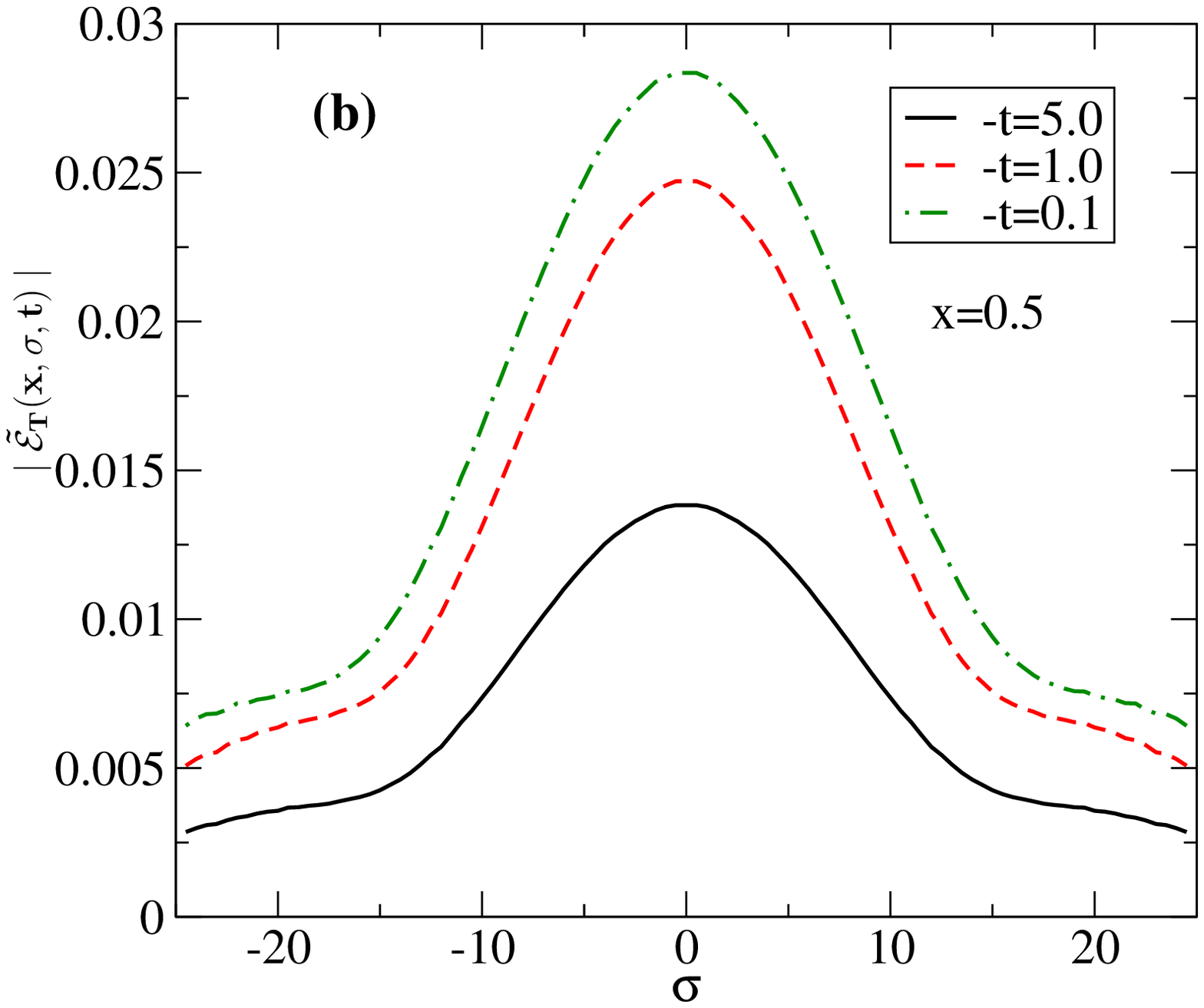}
\includegraphics[width=5cm,height=5cm,clip]{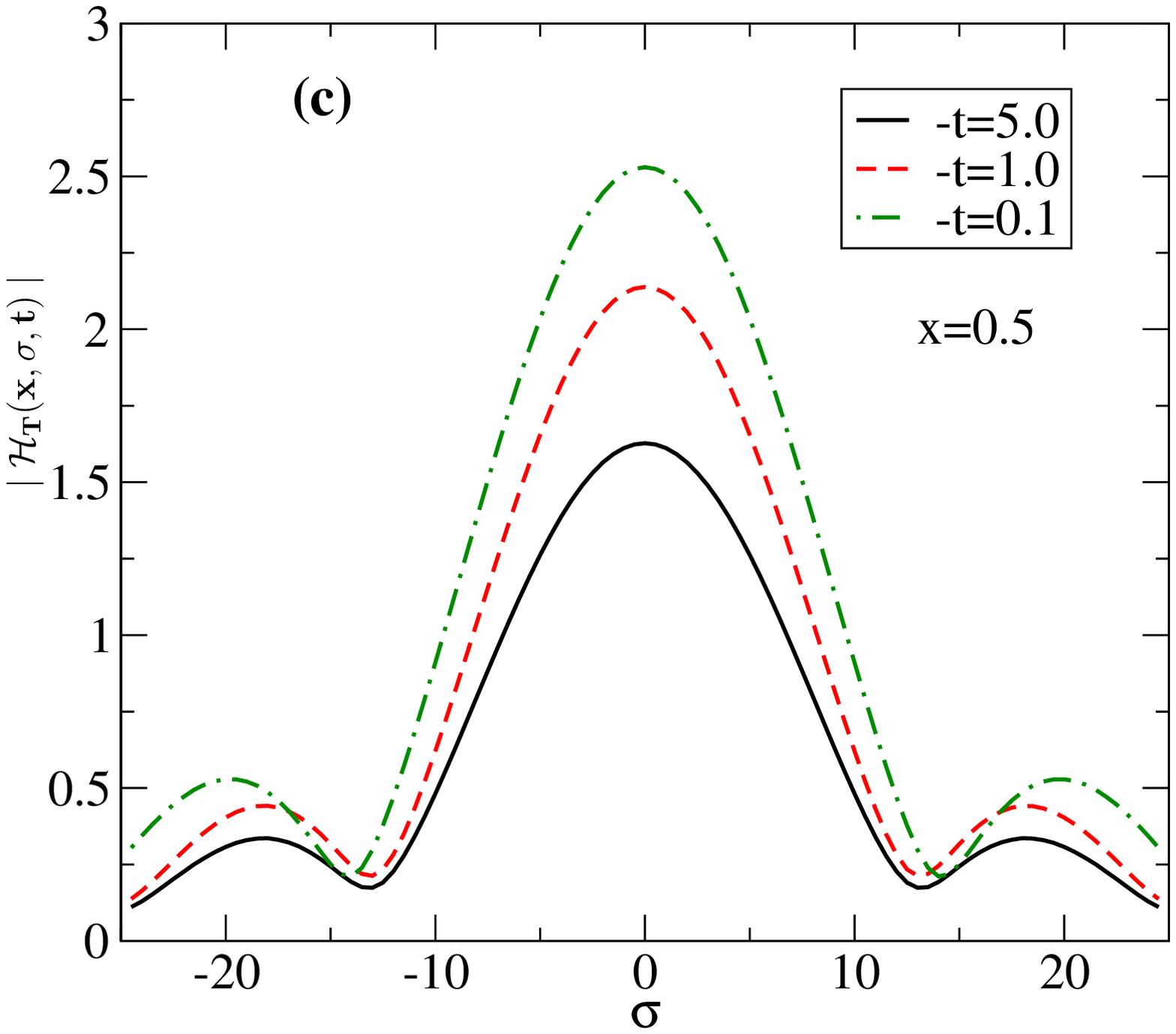}
\caption{\label{fig6} (Color online) Fourier spectrum of the chiral-odd
GPDs vs. $\sigma$  for fixed $x$  and different values of $-t$ in
${\mathrm{MeV}^2}$.}
\end{figure}
We define a boost invariant impact parameter conjugate to the  longitudinal
momentum transfer  as
$\sigma={1\over 2}b^-P^+$ \cite{hadron_optics}.  The chiral odd GPD $E_T$
in longitudinal position space is given by :
\be
{\cal E}_T(x,\sigma,t)&=& {1\over 2\pi}\int_0^{\zeta_f}d\zeta
e^{i{1\over 2}P^+\zeta b^-} E_T(x,\zeta,t)\nonumber \\
&=&{1\over 2 \pi}\int_0^{\zeta_f} d\zeta e^{i\sigma\zeta} E_T(x,\zeta,t).
\ee
Similarly one can obtain ${\cal H}_T(x,\sigma,t)$ and $\tilde {\cal E}_T
(x,\sigma,t)$ as well. Fig. \ref{fig6} shows the plots of the Fourier
spectrum  chiral odd GPDs in
longitudinal position space  as a function of $\sigma$ for fixed $x=0.5$ and
different values of $t$. Both ${\cal E}_T(x,\sigma,t)$ and
${\cal H}_T(x,\sigma,t)$ show diffraction pattern as observed for the DVCS
amplitude in \cite{hadron_optics}; the minima occur at the sames values of
$\sigma$ in both cases.  However $\tilde {\cal E}_T(x,\sigma,t)$
does not show diffraction pattern.
This is due to the distinctively different behaviour of
$\tilde E_T(x,\zeta,t)$ with $\zeta$ compared to that of  $E_T(x,\zeta,t)$
and
$H_T(x,\zeta,t)$. $\tilde E_T(x,\zeta,t)$ rises smoothly from zero and
has no flat plateau in $\zeta$ and thus does not exhibit any diffraction
pattern when Fourier transformed with respect to $\zeta$.
The position of first minima in  Fig.\ref{fig6} is determined by $\zeta_f$.
For $-t=5.0$ and $1.0$, $\zeta_f \approx x=0.5$
 and thus the first minimum appears at the same position while for
$-t=0.1$,
$\zeta_f =\zeta_{max}\approx 0.45$ and the minimum appears  slightly
shifted.
This is analogous to the single slit optical diffraction pattern.
$\zeta_f$ here plays the role of the slit width.
Since the positions of the minima(measured from the center of
the diffraction pattern) are inversely proportional to the slit width,
the minima
move away from the center as the slit width (i.e., $\zeta_f$) decreases.
The  optical analogy of the diffraction pattern in $\sigma$ space has been
discussed in \cite{hadron_optics} in the context of DVCS amplitudes.
\section{Conclusion}
In this work, we studied the chiral-odd GPDs in transverse and
longitudinal position space. Working in light-front gauge, we used
overlap formulas for the chiral odd GPDs in terms of proton
light-front wave functions  in the DGLAP region. 
We used a self consistent relativistic two-body model, namely the quantum
fluctuation  of an electron at one loop in QED. We used its most general
form \cite{drell}, where we have a different mass for the external electron
and different masses for the internal electron and photon.
The impact parameter space representations are obtained by taking Fourier
transform of the GPDs with respect to the transverse momentum transfer. 
When $\zeta$ is non-zero, the initial and final proton are displaced in the
impact parameter space relative to each other by an amount proportional to
$\zeta$. As this is the region probed by most experiments, it is of interest
to investigate this. By taking a Fourier transform with respect to $\zeta$
we presented the GPDs in the boost invariant longitudinal position space
variable
$\sigma$. $H_T$ and $E_T$ show diffraction pattern in $\sigma$ space.


\begin{theacknowledgments}
AM thanks the organizers of Spin 2008 for invitation and support and MHRD for
support.
\end{theacknowledgments}



\end{document}